\documentstyle[osa,manuscript]{revtex}

\begin{document}
\title{Magnetic field dependence of the critical current of tridimensional YBa$_{2}$%
Cu$_{3}$O$_{7-\delta }$ Josephson junction arrays}
\author{W. A. C. Passos*, P. N. Lisboa-Filho and W. A. Ortiz}
\address{Grupo de Supercondutividade e Magnetismo\\
Departamento de F\'{i}sica, Universidade Federal de S\~{a}o Carlos\\
Caixa Postal 676 - 13565-905 S\~{a}o Carlos, SP, Brazil}
\maketitle

\begin{abstract}
In this paper we determine the magnetic field dependence of the critical
current of a tridimensional disordered Josephson junction array (3D-DJJA). A
contactless configuration, employing measurements of the AC-susceptibility,
is used to evaluate the average critical current of an array of YBa$_{2}$Cu$%
_{3}$O$_{7-\delta }$. The critical field necessary to switch off
supercurrents through the weak links at the working temperature is also
obtained.
\end{abstract}

%
%
%

%
\begin{verbatim}
*Corresponding Author: pwac@df.ufscar.br
\end{verbatim}

\bigskip

Tridimensional disorderd Josephson junction arrays (3D-DJJA) can be produced
in a controlled manner, as we have recently demonstrated\cite{1passos}.
Arrays of conventional superconductors (LTS), fabricated from classified
powder, and from ceramic high-temperature (HTS) superconductors, prepared
using a Sol-Gel route, have proved to exhibit all relevant signatures of a
JJA. Those include the typical Fraunhofer dependence of the critical current
with the applied magnetic field, $J_{c}(H)$, the magnetic remanence, and the
Wohlleben effect (WE)\cite{1passos}$^{-}$\cite{3passos}. In fact, these
features are controlled by the McCumber parameter, $\beta _{L}=2\pi
J_{c}(T)L/\phi _{0}$ -where $L$ is the plaquette inductance, and appear only
if $\beta _{L}>>1$\cite{4araujo}. This strong agreement brings us to believe
that average $\beta _{L}$ is in the range, in the case of our samples. But,
it is difucult to infer the actual value of $\beta _{L}$ because the
irregularity of our 3D-DJJA.

Even though behaving as genuine Josephson junction arrays, these 3D-DJJAs
can also be envisaged as especially assembled specimens of granular
superconductors, so that their transport and magnetic properties can be
studied by the commonly employed approaches based on critical state models 
\cite{5fietz}$^{,}$\cite{6chen}. A particularly effective experiment to
determine the average critical current density, $<J_{c}>$, of multileveled
granular structures\cite{7araujo}$^{-}$\cite{9araujo}, consists of measuring
the isothermal AC susceptibility, $\chi _{AC}(h)$, as a function of the
magnetic field amplitude, $h$. The imaginary component, $\chi $'', related
to energy losses, peaks at a field $h_{p}$ which, in turn, is an indirect
measure\cite{7araujo} of $<J_{c}>$, i.e., $h_{p}=a<J_{c}>$ for a sample of
cylindrical shape of radius $a$. This contribution contains results of a
systematic study of the isothermal $\chi _{AC}(h)$ for a 3D-DJJA fabricated
from YBa$_{2}$Cu$_{3}$O$_{7-\delta }$. The experiments were carried out for
different values of the applied magnetic field $H$, and the intergranular
peak of $\chi $'', at $h=h_{p}$, was used to determine the average $%
<J_{c}(H)>$. The present approach is a practical alternative - employing $%
\chi $''$(H)$ instead of $\chi $'$(H)$ - to the method used by
Araujo-Moreira and coworkers\cite{4araujo} to determine $J_{c}(H)$ for JJAs.
It proved to be particularly useful for high temperatures in the $h$%
-parallel-to-$H$ configuration, where the static flux imposed to the array
affects $\chi $'$(H)$, as a background to be subtracted, but not $\chi $''$%
(H)$.

Granular YBCO material used to fabricate the arrays was prepared employing a
modified method of polymeric precursors\cite{10kakihana}. This route
consists of mixing oxides and carbonates in stoichiometric amounts dissolved
in HNO$_{3}$, and then to an aqueous citric acid solution. A metallic
citrate solution is then formed, to which ethylene glycol is added,
resulting in a blue solution which was neutralized to pH$\symbol{126}$7 with
ethylenediamine. This solution was turned into a gel and subsequently
decomposed to a solid by heating at 400 $^{o}$C. The sample was heat-treated
at 850 $^{o}$C for 12 h in air with several intermediary grindings, in order
to prevent undesirable phase formations. Then, it was pressed into a pellet
using controlled uniaxial (5,000 kgf/cm$^{2}$) pressure and sintered at 9500 
$^{o}$C for 6 h in O$_{2}$. This pellet is a 3D-DJJA, in which the junctions
are weakly coupled grains with 5 $\mu $m of average diameter, i.e.,
weak-links (WLs) formed by sandwiches of YBCO grains and intergrain
material. As a consequence of the uniaxial pressure, samples produced in
this way are anisotropic, a feature that can be either enhanced, by using
higher pressures, or reduced, by applying isostatic pressures. Also, thermal
treatment plays a fundamental role on creation and control of WLs and
anisotropy, as is thoroughly discussed elsewhere\cite{1passos}.

AC-susceptibility measurements, $\chi _{AC}(h)$, were carried out using the
AC-module of a Quantum Design SQUID magnetometer, for the excitation field $%
0.01\leq h\leq 3.8$ Oe and frequency of 100 Hz, at temperatures ranging from 
$T=2$ K up to 100 K. In this paper we focus on the field dependence for $T=78
$ K, close enough to $T_{c}$ to ensure that hp is below 3.8 Oe, the upper AC
field achievable in the experimental apparatus employed. Fig.1 shows the
real ($\chi $') and imaginary ($\chi $'') parts of $\chi _{AC}(h)$ for some
values of $H$. As mentioned above, $h_{p}$ is an indirect measure of $<J_{c}>
$ which, in the present case, is the average critical current density of the
3D-DJJA\cite{7araujo}$^{-}$\cite{9araujo}. In Fig.2, the average $J_{c}(H)$
for $T=78$ K is shown. The line connecting the experimental points for the
average $J_{c}(H)$ is a fit of the form $%
J_{c}(H,T)=J_{c0}[A+B(H^{2}+C^{2})^{-1}](1-T/Tc)^{2.38}$, introduced by
Wright and coworkers\cite{11wright}$^{,}$\cite{12wright} for a matrix formed
by grains linked by Josephson couplings. The fitting parameters appear in
the inset table. It is worth mentioning that $T_{c}$ is not the grain's
critical temperature, but the array's instead, reported previously for this
same sample as $T^{\ast }=83$ K. Also, from the values arising from the
fitting one concludes that a critical field $H_{c}=35$ Oe is sufficient to
suppress superconductivity through the 3D-DJJA at any temperature, as the
superconducting current across the weak-links vanishes for fields above $%
H_{c}$.

To summarize, we recall that, besides current transport across the sample,
which would be infeasible here, two other methods could be employed to
determine the magnetic field dependence of the critical current of 3D-DJJAs:
(i) measure of $\chi $'$(H)$ with $H$ parallel to the plaquettes\cite
{4araujo}; and (ii) measure of $\chi $'$(H)$ with $H$ perperdicular to the
plaquettes\cite{1passos}$^{-}$\cite{3passos}. Method (i.) would require
static and sinusoidal fields perpendicular to each other, what is
unattainable in our experimental apparatus, as well as in most of the
existing laboratory setups. Method (ii.) imposes static flux into the array,
affecting $\chi $'$(H)$, so that a background has to be subtracted. Using
the present method, however, the average critical current density is
obtained in a straightforward manner, since $\chi $''$(H)$ is not affected
by the static flux.

Financial support was partially provided by Brazilian agencies FAPESP,
CAPES, PRONEX and CNPq.

\begin{figure}[tbp]
\caption{Real ($\protect\chi $') and imaginary ($\protect\chi $'') parts of $%
\protect\chi _{AC}(h)$ for three values of the static applied field, $H$ =
5, 10 and 15 Oe. The excitation field at which c'' peaks is an indirect
measure of the average critical current density of the array.}
\end{figure}

\begin{figure}[tbp]
\caption{Average $J_{c}(H)$ for $T$ = 78 K. The line connecting the
experimental points is a fit of the form $%
J_{c}(H,T)=J_{c0}[A+B(H^{2}+C^{2})^{-1}](1-T/Tc)^{2.38}$. $T_{c}$ is the
array critical temperature. The appropriate combination of the fitting
parameters lead to $H_{c}$ = 35 Oe, above which the WLs in the array do not
superconduct at this temperature.}
\end{figure}

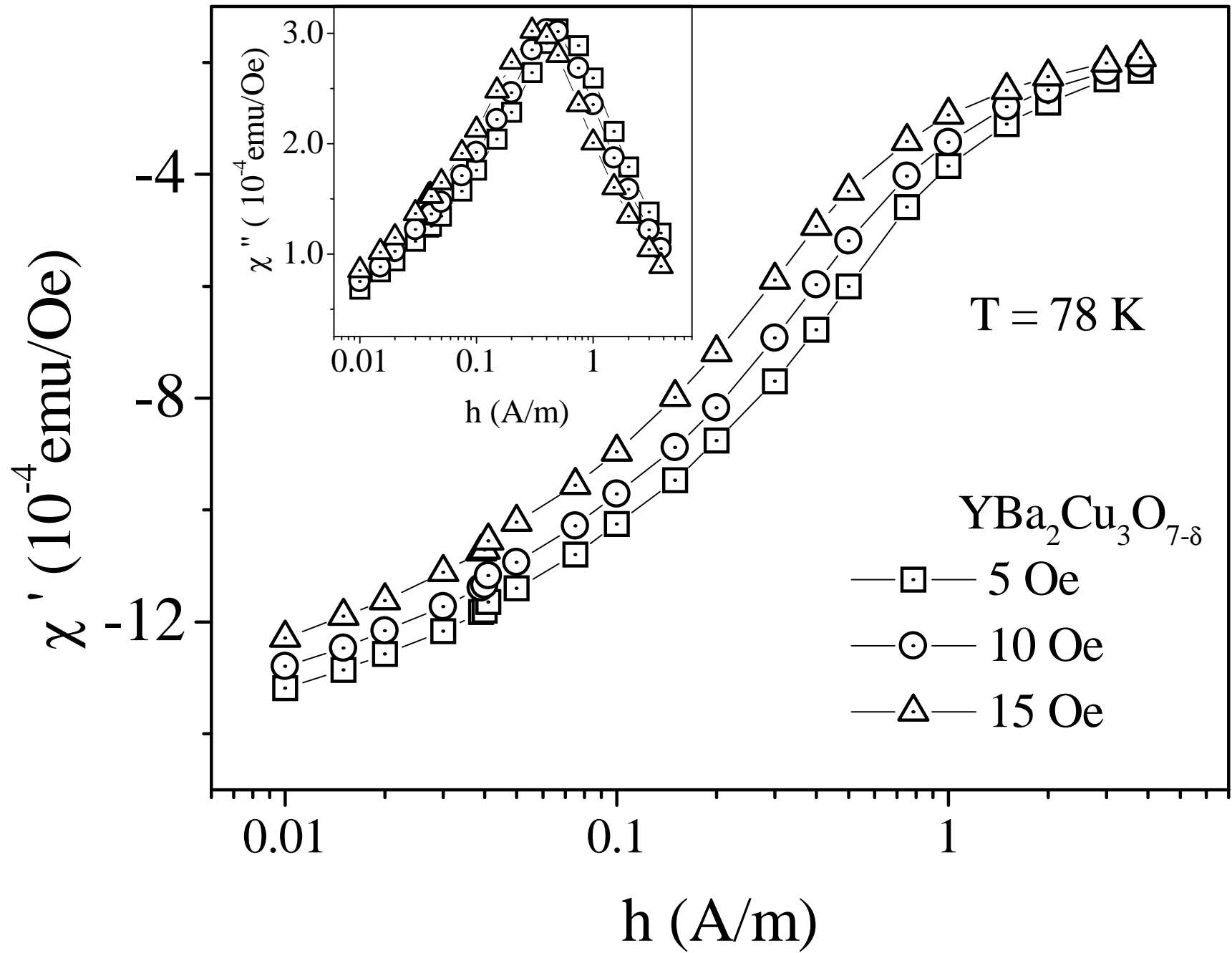
\end{document}